\documentclass[twocolumn,showpacs,preprintnumbers,amsmath,amssymb]{revtex4}

\usepackage{graphicx}

\newcommand{\correct}{}

\begin{document}


\title{Fidelity decay in trapped Bose-Einstein condensates}
\author{G. Manfredi}
\email{Giovanni.Manfredi@ipcms.u-strasbg.fr}
\author{P.-A. Hervieux}
\affiliation{Institut de Physique et Chimie des Mat{\'e}riaux, CNRS
and Universit{\'e} Louis Pasteur, BP 43, F-67034 Strasbourg, France}

\date{\today}
\begin{abstract}
The quantum coherence of a Bose-Einstein condensate is studied
using the concept of quantum fidelity (Loschmidt echo). The
condensate is confined in an elongated anharmonic trap and
subjected to a small random potential such as that created by a
laser speckle. Numerical experiments show that the quantum
fidelity stays constant until a critical time, after which it
drops abruptly over a single trap oscillation period. The critical
time depends logarithmically on the number of condensed atoms and
on the perturbation amplitude. This behavior may be observable by
measuring the interference fringes of two condensates evolving in
slightly different potentials.
\end{abstract}

\pacs{03.75.Gg, 05.45.Mt}


\maketitle

{\it Introduction}.--- Ultracold atom gases have been at the
center of intensive investigations since the first realization of
an atomic Bose-Einstein condensate (BEC) in the mid-1990s.
Applications of BECs include the possibility of revisiting
standard problems of condensed-matter physics \cite{Anglin}, by
making use of periodic optical lattices that mimic the ionic
lattice in solid-state systems. More recently, BECs have been used
to study quantum transport in disordered systems (another
long-standing problem in condensed matter physics), by using a
laser speckle to create a disordered potential. For instance, it
was shown that the free expansion of a BEC is restrained, or even
completely suppressed, in the presence of a disordered potential
\cite{Fort,Clement} (see also \cite{Modugno} for a theoretical
analysis), an effect akin to Anderson localization in solids
\cite{Paul}.

These problems are often approached by studying the interference
pattern of two or more BECs that are released from the optical
trap, expand freely, and eventually interact with each other
\cite{Andrews,Kasevich,Greiner,Shin}. Experiments show
high-contrast matter-wave interference fringes, thus revealing the
coherent nature of Bose-Einstein condensates. Surprisingly, recent
experiments have shown high-contrast fringes even for well
separated BECs, whose phases are totally uncorrelated
\cite{Hadzi}.

Because the interference pattern depends on the phases of the
condensates, the fringes should be sensitive to perturbations that
strongly affect the phase, but weakly affect the motion.
Therefore, when two condensates are subjected to a random
potential (such as that generated by a laser speckle \cite{Lye,
Clement2}) before interfering, we expect a reduction in the
contrast of the interference fringes, which should depend both on
the amplitude of the random potential and on the time during which
it has been in contact with the condensates. The purpose of this
Letter is to propose a theoretical procedure to quantify this loss
of coherence and to suggest a possible experimental realization.

{\correct In order to estimate the coherence and stability of a
quantum system \cite{peres}, one can compare the evolution of the
same initial condition in two slightly different Hamiltonians,}
$H_1=H_0+\delta H_1$ and $H_2=H_0+\delta H_2$, where $H_0$ is the
unperturbed Hamiltonian, and $\delta H_{1,2}$ are small
perturbations characterized by the same amplitude, same wavelength
spectrum, but different phases. The quantum fidelity at time $t$
is then defined as the square of the scalar product of the
wavefunctions evolving with $H_1$ and $H_2$ respectively:
$F(t) = ~\vline \langle \psi_{H_1}(t) \vline ~\psi_{H_2}(t)
\rangle \vline^2$.
This procedure is sometimes referred to as the `Loschmidt echo',
as it is equivalent to evolving the system forward in time with
$H_1$, then backward with $H_2$, and using the fidelity to check
the accuracy of the time-reversal.

Virtually all theoretical investigations of the Loschmidt echo
consider one-particle systems evolving in a given (usually
chaotic) Hamiltonian. Several regimes have been described in the
past. For perturbations that are classically weak but
quantum-mechanically strong, the fidelity decay is exponential,
with a rate independent on the perturbation and given by the
classical Lyapunov exponent of the unperturbed system
\cite{jalabert}. This behavior has been confirmed by numerical
simulations \cite{jacquod,cucchietti}. For weaker perturbations,
the decay rate is still exponential, but perturbation-dependent
(Fermi golden rule regime). For still weaker perturbations, the
decay is Gaussian (perturbative regime) \cite{jacquod}. For
integrable systems, other types of decay (notably algebraic) have
been observed \cite{benenti}. Finally, a perturbation-independent
regime, though with Gaussian decay, was also observed in
experiments \cite{Pastawski}.

In a previous work \cite{Manfredi}, we have applied the concept of
Loschmidt echo to a system of many electrons interacting through
their self-consistent electric field. The numerical results showed
that the quantum fidelity remains equal to unity until a critical
time, then drops suddenly to much lower values. A similar result
was also obtained for a classical system of colliding hard spheres
\cite{Pinto}. This effect is probably related to the nonlinearity
introduced by the interactions between particles. Therefore, BECs
should constitute an ideal arena to determine whether such
behavior is typical of many-body quantum systems.

{\it Model}.---The dynamics of a BEC is accurately described, in
the mean-field approximation, by the Gross-Pitaevskii equation
(GPE). We considered a cigar-shaped condensate, where the
transverse frequency of the confining potential is much larger
then the longitudinal frequency, $\omega_\perp \gg \omega_z$. In
this case, a one-dimensional (1D) approximation can be used, and
the GPE reads as:
\begin{equation}
i\hbar\frac{\partial\psi}{\partial\,t} = -
\,\frac{\hbar^2}{2m}\frac{\partial^{2}\psi}{\partial\,z^2} +
V(z)\psi + g_{1D}N_A|\psi|^2\psi \equiv H_0 \psi \, . \label{GP}
\end{equation}
Here, $\int_{-\infty}^{\infty} |\psi|^2 dz=1$, $N_A$ is the number
of condensed atoms, $g_{1D} = 2a\hbar\omega_\perp$ is the 1D
effective coupling constant, and $a$ is the 3D scattering length.
The confining potential contains a small quartic component, which
can be realized optically \cite{Bretin}, $V(z) =
\frac{1}{2}m\omega_z^2( z^2 + Kz^4/L_{\rm ho}^2)$, where $L_{\rm
ho}=(\hbar/m\omega_z)^{1/2}$ is the harmonic oscillator length.

We choose the parameters of the experiment described in Ref.
\cite{Fort}, where $N_A = 10^5$ atoms of $^{87}\rm Rb$ ($a =
5.7~\rm nm$) are confined in a quasi-1D trap with $\omega_z/2\pi =
24.7~\rm Hz$ and $\omega_\perp/2\pi = 293~\rm Hz$. In the
simulations, we normalize time to $\omega_z^{-1}$, space to
$L_{\rm ho} = 2.16~\rm \mu m$, and energies to $\hbar\omega_z$.
The dimensionless 1D coupling constant is then: $\hat{g}_{1D} =
g_{1D}/(L_{\rm ho}\hbar\omega_z)=0.063$. The quartic coefficient,
which will be crucial to excite a sufficiently complex nonlinear
dynamics, is taken to be $K=0.05$. With the above parameters, the
half-length of the condensate is roughly $24~\rm \mu m$.

The random perturbation $\delta H$ can be realized in practice
using a laser speckle \cite{Lye,Clement2}. In our simulations, we
model the random potential through the sum of a large number of
uncorrelated waves: $\delta H/\hbar\omega_z = \epsilon
\sum_{j=N_{\rm min}}^{N_{\rm max}} \cos(2\pi z/\lambda_j
+\alpha_j)$, where $\epsilon$ is the amplitude of the
perturbation, $\lambda_j$'s are the wavelengths, and $\alpha_j$'s
are random phases. The smallest wavelength present in the random
potential is $\lambda_{\rm min} = 2L_{\rm ho} = 4.32~\rm \mu m$,
which is consistent with the experimental correlation length,
$\sigma_z = 5~\rm \mu m$ \cite{Fort}. The wavelength spectrum of
the perturbation (i.e. the values of $N_{\rm min}$ and $N_{\rm
max}$) affects only weakly the behavior of the fidelity:
therefore, we will focus our analysis on the dependence of the
fidelity on the amplitude $\epsilon$.

In order to compute the quantum fidelity, we proceed as follows:
(i) first, we prepare the condensate in its ground state without
perturbation; (ii) then, we suddenly displace the anharmonic trap
by a distance $\Delta z$ (a few micrometers); (iii) finally, we
solve numerically the time-dependent GPE (\ref{GP}) with the
perturbed Hamiltonian $H_0+\delta H$. Step (iii) is performed for
$N=11$ uncorrelated realizations of the random potential, thus
yielding $N$ evolutions of the wavefunction, $\psi_j(t)$. We then
use all possible combinations to compute the partial fidelities
$F_{ij}(t) = ~\vline \langle \psi_{i}(t) \vline ~\psi_{j}(t)
\rangle \vline^2$. There are of course $M=N!/(N-2)!2!=55$
independent combinations, which are finally averaged to obtain the
quantum fidelity, $F(t)=\frac{1}{M}\sum_{j=1}^M F_{ij}(t)$. This
averaging procedure allowed us to reduce considerably the level of
statistical fluctuations.

\begin{figure}[htb]
\includegraphics[height=4cm]{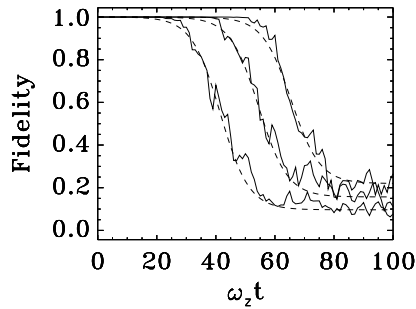}
\caption{\label{fig:fig1} Fidelity decay for $\Delta z = 3L_{\rm
ho}$, $N_A=10^5$, and three values of the perturbation: $\epsilon
=10^{-3}$, $\epsilon =10^{-5}$, $\epsilon =10^{-7}$ (the critical
time increases with decreasing perturbation). The dashed lines are
fits obtained using Eq. (\ref{eq:fit}) with $\omega_z T = 4.86$.}
\end{figure}

{\it Results}.--- Our numerical results showed an unusual behavior
for the quantum fidelity, which stays equal to unity until a
critical time $\tau_C$, and then drops rapidly to small values
(Fig. 1). The critical time is defined as the time at which the
fidelity has dropped to 60\% of its maximum value, i.e.
$F(\tau_C)=0.6$. Interestingly, the fidelity can be nicely fit by
a Fermi-like curve
\begin{equation}
f(t) =
(1-f_\infty)\left[1+\exp\left(\frac{t-\tau_C}{T}\right)\right]^{-1}
+ f_\infty.\label{eq:fit}
\end{equation}
Equation (\ref{eq:fit}) reveals the presence of two distinct time
scales: (i) the critical time $\tau_C$ and (ii) $T \ll \tau_C$,
which measures the rapidity of the fidelity decay. The parameter
$f_\infty$ simply reflects the fact that the fidelity cannot decay
to zero, because the system is confined in a finite region in
space. In Fig. 1, $\omega_z T = 4.86$ is the same for all three
cases and is equal to the oscillation period of a particle trapped
in the anharmonic potential $V(z)$, for an initial condition $z(0)
= \Delta z$, $\dot z(0)=0$. This means that {\em the fidelity
decay occurs over one single oscillation period}.

\begin{figure}[htb]
\includegraphics[height=4cm]{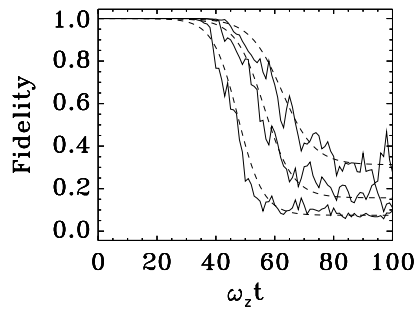}
\caption{\label{fig:fig2} Fidelity decay for $N_A=10^5$, $\epsilon
=10^{-5}$, and different values of the displacement $\Delta z =
2L_{\rm ho}$, $\Delta z = 3L_{\rm ho}$, and $\Delta z = 4L_{\rm
ho}$. The steeper curves correspond to larger values of $\Delta
z$. The dashed lines are fits obtained using Eq. (\ref{eq:fit}),
with $\omega_z T = 4.27$ ($\Delta z = 4L_{\rm ho}$), $\omega_z T =
4.86$ ($\Delta z = 3L_{\rm ho}$), and $\omega_z T = 5.50$ ($\Delta
z = 2L_{\rm ho}$).}
\end{figure}

This behavior was confirmed by investigating the dependence of the
quantum fidelity on the initial displacement $\Delta z$ (Fig. 2).
For each value of $\Delta z$, the parameter $T$ appearing in Eq.
(\ref{eq:fit}) is taken to be equal to the corresponding
oscillation period in the anharmonic potential $V(z)$. The
oscillation period decreases with increasing energy (and thus with
increasing $\Delta z$) and indeed the fidelity drop becomes
steeper for larger displacements. The critical time also slightly
increases with decreasing displacement and goes to infinity for
$\Delta z \to 0$. This presumably happens because the dynamics of
the condensate becomes too regular when the confinement is
harmonic.

\begin{figure}[htb]
\includegraphics[height=4cm]{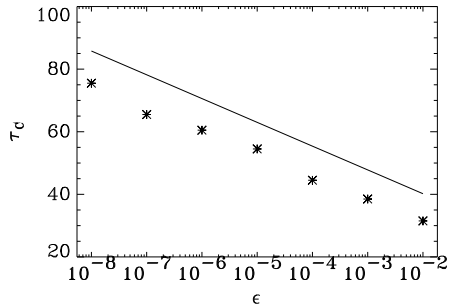}
\caption{\label{fig:fig3} Critical time $\tau_C$ (in units of
$\omega_z^{-1}$) versus perturbation amplitude $\epsilon$, for
$\Delta z = 3L_{\rm ho}$ and $N_A = 10^5$. The solid line
represents the curve $\tau_C \sim -t_0 \ln \epsilon$, with
$\omega_z t_0 = 3.3$.}
\end{figure}

Figure 3 shows that $\tau_C$ depends logarithmically on the
perturbation amplitude, i.e. $\tau_C \sim -t_0 \ln \epsilon$, with
$\omega_z t_0 \simeq 3.3$ (this is the straight line depicted in
Fig. 3). This is similar to what was obtained for another
self-consistent model \cite{Manfredi}, suggesting that such
behavior is generic for $N$-body systems, at least in the mean
field approximation. The critical time also depends on the number
of condensed atoms $N_A$. By decreasing $N_A$, $\tau_C$ becomes
considerably longer (Fig. 4), and for $N_A \to 0$ (i.e. for the
linear Schr{\"o}dinger equation) we have that $\tau_C \to \infty$. For
sufficiently large condensates ($N_A \ge 2\times 10^4$ for the
case of Fig. 4), the critical time depends logarithmically on the
number of atoms.

{\correct Finally, the collapse of the quantum fidelity is clearly
linked to the phases of the wavefunctions. Indeed, by defining an
`amplitude fidelity' $F_a(t) = ~\left(\int|\psi_{H_1}\psi_{H_2}|dx
\right)^2$ (which neglects information on the phases), we have
verified that $F_a(t)$ shows no sign of a sudden collapse when the
ordinary fidelity drops.}

\begin{figure}[htb]
\includegraphics[height=4cm]{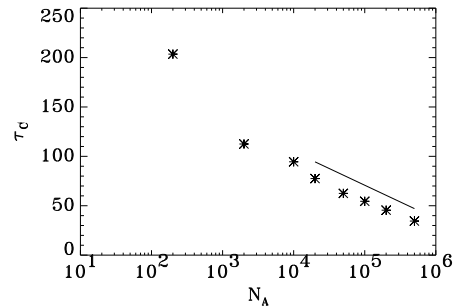}
\caption{Critical time $\tau_C$ (in units of $\omega_z^{-1}$)
versus number of atoms $N_A$ in the condensate, for $\Delta z =
3L_{\rm ho}$ and $\epsilon = 10^{-5}$. The straight line is a
guide to the eye.} \label{fig:fig4}
\end{figure}

{\it Discussion}.---{\correct We have shown that the evolutions of
two BECs in slightly different Hamiltonians diverge suddenly after
a critical time $\tau_C$. The interaction between the atoms is
obviously a vital ingredient, as $\tau_C \to \infty$ when the
coupling constant vanish, i.e. for the linear Schr{\"o}dinger
equation. The crucial point is that, for the GPE, the unperturbed
Hamiltonian $H_0$ depends on the wave function.} When the
perturbation induces a small change in $\psi$, $H_0$ is itself
modified, which in turns affects $\psi$, and so on. Thanks to such
a nonlinear loop, the perturbed and unperturbed solutions can
diverge very fast. In contrast, for the single-particle dynamics
$H_0$ is fixed, and the solutions only diverge because of the
perturbation $\delta H$. Changes in $\psi$ add incrementally to
each other, but cannot trigger the nonlinear loop observed in the
GP simulations \cite{foot}.

{\correct This behavior is clearly linked to the phases of the
wavefunctions and could be tested experimentally by studying the
effect of a random potential on the interference pattern of two
condensates \cite{Andrews,Kasevich,Greiner,Shin}.} A possible
experiment could be performed as follows (see Fig. 5). First, a
BEC is created in a single-well trap; then, the trap is deformed
into a double-well potential \cite{Shin,Hansel}, with the barrier
between the wells sufficiently high that the two condensates
cannot tunnel through it. The condensates are left in the double
trap for a time long enough to reach their ground state. A laser
speckle is then used to create a small random potential of
amplitude $\epsilon$ and correlation length $\sigma_z$. If
$\sigma_z \ll d$, where $d$ is the distance between the two BECs,
each condensate is subjected to a different random potential with
the same statistical properties.

In order to excite the dynamics, the total double-well trap is
suddenly shifted by a distance $\Delta z$ of the order of a few
micrometers. The BECs evolve in their perturbed trap for a certain
time $t$, after which both the trap and the random potential are
switched off, so that the condensates can overlap and interfere.
We predict that the contrast of the interference fringes will
depend on the time $t$ and on the perturbation $\epsilon$, in a
manner analogous to the quantum fidelity: if $t<\tau_C(\epsilon)$,
the contrast should be large, whereas it should drop significantly
for times larger than $\tau_C$. Performing several experiments
with different perturbation amplitudes and different numbers of
condensed atoms should allow one to reproduce qualitatively the
logarithmic scalings of Figs. 3 and 4. Accurate time-resolved
measurements might even reproduce the fidelity drop time $T$.

\begin{figure}[hbt]
\includegraphics[height=5cm]{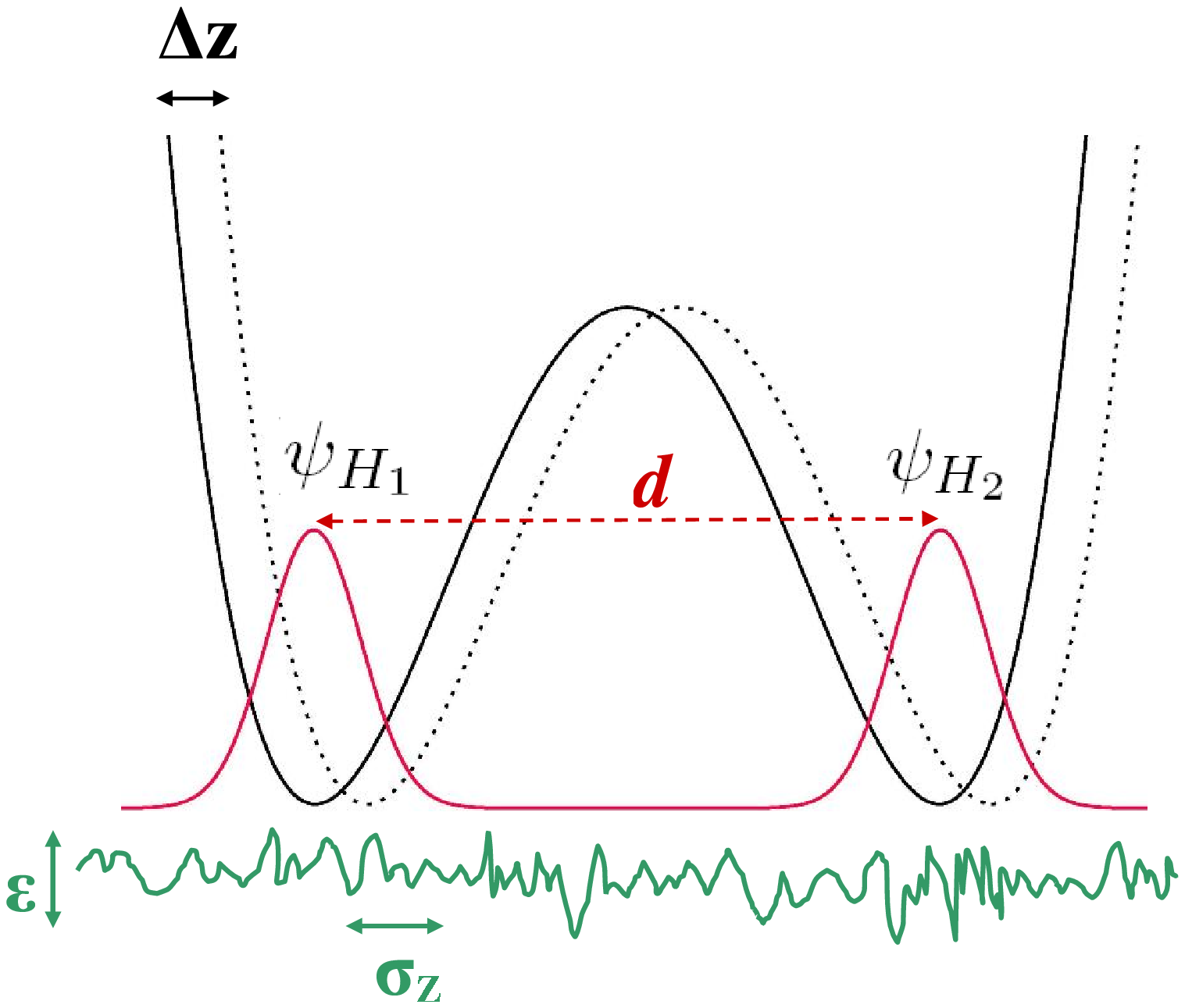}
\caption{Two BECs ($\psi_{H_1}$  and $\psi_{H_2}$) are created in
a double-well trap. A random potential with amplitude $\epsilon$
and correlation length $\sigma_z$ is superimposed to the confining
potential. In order to excite the nonlinear dynamics, the trap is
suddenly shifted by a distance $\Delta z$.} \label{fig:fig5}
\end{figure}

These results, together with those obtained for an electron gas at
low temperature \cite{Manfredi}, suggest that many-particle
systems display a generic sudden decay of the quantum fidelity.
Thanks to the ease with which ultracold atom gases can be created
and manipulated in the laboratory, BECs should constitute an ideal
arena to test these predictions experimentally.

We thank C. Fort for providing the details of the experiment
described in Ref. \cite{Fort}. We also thank R. Jalabert, J.
L{\'e}onard, and H. Pastawski for several useful suggestions.


\begin{references}
\bibitem{foot} As the GPE is intrinsically nonlinear, one could, in principle, be
able to merely perturb the initial quantum state (without doing
anything to the Hamiltonian) in order to see some nontrivial decay
of the corresponding overlap function. It will be interesting to
investigate this alternative approach in future studies.
\end{references}
\end{document}